\documentclass[prb,twocolumn,a4paper,showpacs,aps]{revtex4}
\pdfoutput=1
\usepackage{amsmath, amssymb}
\usepackage{amsfonts}
\usepackage{graphicx}
\usepackage{color}

\newcommand{\nn}{\nonumber}
\newcommand{\g}{\gamma}

\newcommand{\eff}{\mathrm{eff}}

\begin{document}

\title{Nonequilibrium transport via spin-induced sub-gap states in
  superconductor/quantum dot/normal metal cotunnel junctions}
\author{V.~Koerting$^{1,2}$}
\email[author to whom correspondence should be addressed:\ ]
{koerting@nbi.dk}
\author{B.~M.~Andersen$^1$}
\author{K.~Flensberg$^1$}
\author{J.~Paaske$^1$}
\affiliation{$^1$Nano-Science Center, Niels Bohr Institute, University
of Copenhagen, Universitetsparken 5, DK-2100~Copenhagen \O , Denmark
\\
$^2$Niels Bohr International Academy, Niels Bohr Institute, University
of Copenhagen, Blegdamsvej 17, DK-2100~Copenhagen \O , Denmark}

\begin{abstract}
We study low-temperature transport through a Coulomb blockaded
quantum dot (QD) contacted by a normal (N), and a
superconducting (S) electrode. Within an effective cotunneling
model the conduction electron self energy is calculated to
leading order in the cotunneling amplitudes and subsequently
resummed to obtain the nonequilibrium T-matrix, from which we
obtain the nonlinear cotunneling conductance. For \textit{even
occupied} dots the system can be conceived as an effective
S/N-cotunnel junction with subgap transport mediated by Andreev
reflections. The net spin of an \textit{odd occupied} dot,
however, leads to the formation of sub-gap resonances inside the
superconducting gap which gives rise to a characteristic peak-dip structure
in the differential conductance, as observed in recent experiments.
\end{abstract}
\date{\today}
\maketitle

\section{Introduction}

Magnetic impurities in normal metals are known to give rise to
so-called Abrikosov-Suhl
resonances,~\cite{Kondo64,Abrikosov65,Suhl65} which in turn lead to
the celebrated Kondo conductance anomaly observed in normal
(N/barrier/N) tunnel junctions with magnetic impurities in the
barrier,~\cite{Logan64,Wyatt64,Shewchun65,Appelbaum66,Anderson66} as
well as in normal (N/QD/N) cotunnel junctions based on Coulomb
blockaded quantum dots (QD) holding an odd number of
electrons.~\cite{Ng88,Glazman88,Goldhaber98,Cronenwett98,
vanderWiel00,Nygaard00} Magnetic impurities in superconducting
metals, on the other hand, give rise to localized Yu-Shiba-Rusinov
bound states inside the superconducting gap,~\cite{Yu65,Soda67,Shiba68,Rusinov69,
Shiba69} which can be observed by measuring the local density of
states (DOS) in scanning tunneling microscopy
(STM) as sub-gap conductance peaks offset from the gap edge roughly
by the magnitude of the exchange coupling (cf.
Refs.~\onlinecite{Yazdani97,Salkola97,Flatte97,Flatte99,Balatsky06,Moca08} and
references therein).

\begin{figure}[!h]
  \centering
  \includegraphics[width=\columnwidth]{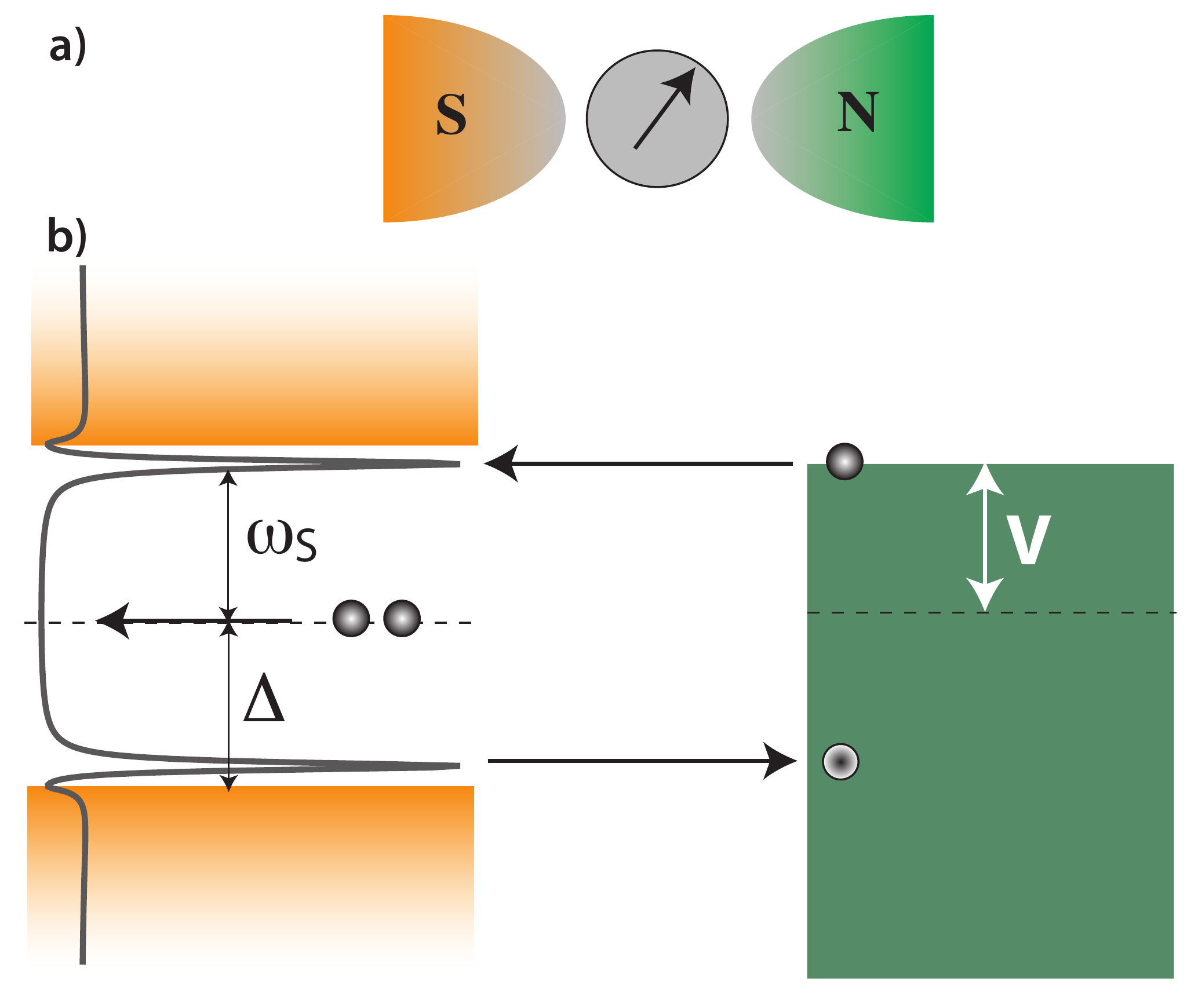}
  \caption{\textbf{a)} Sketch of the S/QD/N cotunnel
  junction comprised by a (gray) quantum dot holding a net spin,
  which is tunnel coupled to an (orange) superconducting and a
  (green) normal metal lead.
  \textbf{b)} Illustration of the basic Andreev reflection
process giving rise to sub-gap transport and how it is enhanced by
the presence of localized resonances induced by the spin on
the quantum dot. The right (green) region illustrates the constant density of
states (DOS) in the normal lead, shifted by the voltage $V$, and the
left (orange) region illustrates the BCS DOS in the superconducting lead. The line is
the local dot-electron DOS, with sub-gap resonances
inside the gap and a pronounced dip at each gap edge.}
  \label{fig:Shiba_Andreev}
\end{figure}

In this paper, we explore the effects of spin-induced bound states in
(S/QD/N) cotunnel junctions based on Coulomb blockaded quantum dots
contacted to one superconducting and to one normal metal lead (cf.
Fig.~\ref{fig:Shiba_Andreev}a). As we shall demonstrate, a coupling
to the normal lead will broaden the localized bound states and the
resulting scattering resonances will be reflected as characteristic
sub-gap \textit{peaks}, accompanied by pronounced \textit{dips} at
the gap edges in the nonlinear conductance. Basically, for a spinful
quantum dot, sub-gap transport via Andreev
reflections~\cite{Andreev64,Blonder82} probes the profile of the sub-gap states rather than simply the bare BCS DOS (cf.~Fig.~\ref{fig:Shiba_Andreev}b).

Transport measurements on such S/QD/N systems in the
cotunneling regime have been already carried
out.~\cite{Graeber04,Deacon10a,Deacon10b} Most recently, Deacon
\textit{et al.}~\cite{Deacon10a,Deacon10b} have indeed observed
sub-gap conductance peaks for odd occupied S/InAs-QD/N devices,
which they interpret as signatures of Andreev energy levels
inside the gap.~\cite{Tanaka07,Bauer07}
Below, we argue that these peaks can be ascribed to Yu-Shiba-Rusinov resonances
forming in a spinful cotunnel junction. This is consistent with the
results of Refs.~\onlinecite{Tanaka07,Bauer07},
but allows for a simpler interpretation and
calculation in terms of the Kondo model rather than the Anderson
model. As we show, the experimental observation of enhanced Andreev
current in spinful dots, conductance dips near the gap edges and
gate-dependence of the sub-gap peak positions can all be explained in
terms of such spin-induced resonances.

A number of works have addressed the problem of an Anderson
impurity coupled to a single superconductor, either by
numerical renormalization group (NRG)
calculations~\cite{Satori92,Bauer07,Hecht08,Lim08} or auxiliary
boson methods,~\cite{Clerk00,Sellier05} and have explored the
intricate competition between Cooper-pairing and local correlations as a function of tunnel coupling $\Gamma$,
charging energy $U$, and superconducting gap $\Delta$. Even a
simple non-interacting ($U=0$) model gives rise to sub-gap
states,~\cite{Beenakker92,Khlus93,Bauer07,Hecht08} which may
affect the nonlinear conductance, and sub-gap states are thus
sustained in many different parameter regimes and possibly with
many different characteristics. The present paper, however,
focuses on quantum dots in the cotunneling regime safely inside
a Coulomb diamond where charge fluctuations are strongly
suppressed. Restricting ourselves to the cotunneling (Kondo)
model, which is far simpler than the Anderson model, we retain
crucial correlation features and, at the same time, we are able
to capture the important physics of spin-induced bound states in a quantum dot setting, even out of equilibrium.

As already mentioned, a finite coupling to the normal metal lead
will change the sub-gap bound states into broadened resonances but for
low enough temperatures, Kondo correlations will become important
and the resonances will be either suppressed, or supplemented
by a Kondo resonance pinned to the normal metal Fermi surface. The
full S/QD/N problem is an inherently complicated
problem,~\cite{Yamada10,Fazio98,Schwab99,Sun01,Cuevas01,Krawiec04,Domanski07,
Domanski08,Tanaka07,Governale08} which we shall not attempt to solve
here. In order to isolate and explore the observable consequences of
the spin-induced sub-gap resonances, we neglect the log-singular terms arising from
Kondo correlations with the normal lead, thus tacitly assuming the
coupling between dot and normal lead to be sufficiently weak such that
the corresponding Kondo temperature, $T_K$, is much smaller than
than either temperature, $T$, or applied bias-voltage, $V$. Staying
with an effective cotunneling model, it would indeed be interesting to
investigate the competition between these sub-gap resonances and Kondo
instabilities in the regime $\Delta\ll T_K$ where nonlinear
conductance has been reported~\cite{Graeber04} to be very
different from the regime $\Delta\gg T_K$, which we study here.

After introducing the model, we explain how to get from lowest order
self-energies to the current via the nonequilibrium T-matrix. Using
this setup, we then start by investigating the case of a spinless
even-occupied quantum dot, for which the effective cotunneling model
takes the form of a simple potential scattering term. The nonlinear
conductance is shown to be the same as for an ordinary S/N
junction,~\cite{Blonder82,Beenakker92,Cuevas96} which is to say that the
spinless quantum dot can be viewed as an effective S/N-cotunnel
junction.

Next, we treat the case studied originally by Yu\cite{Yu65}, Shiba\cite{Shiba68}, and Rusinov\cite{Rusinov69} of a classical spin, modelled by a
\textit{spin-dependent} potential scattering term and determine
again the nonlinear conductance. In this case we can still obtain
exact analytical expressions for the current and comparing the expressions
with the potential scattering case we find that already the
classical spin leads to a \textit{peak} below the gap accompanied by
a square root \textit{dip} at the gap edge (as opposed to the usual
BCS square root divergence).

\begin{figure}[t] \centering
\includegraphics[width=\columnwidth]{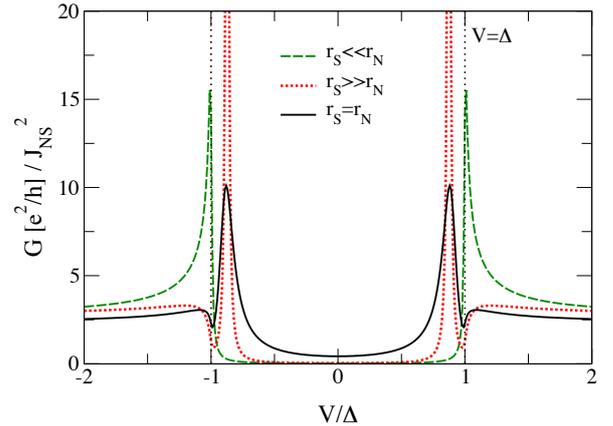}
\caption{Comparison of the nonlinear conductance through an odd
occupied quantum dot holding a net spin for different choices of
coupling asymmetry. The spin exchange interaction is parametrized
by $J_{\alpha\alpha'} = J_0 r_\alpha r_{\alpha'}$. A spin-induced bound state
is seen as a sub-gap resonance for strong coupling to the
superconducting lead $r_S \gg r_N$ (factor of $10$ different) and for
symmetric coupling, though broadened in that case. A dip appears at
the superconducting gap, where one would expect a square root
divergence like in the case of $r_N \gg r_S$, where the normal lead is
stronger coupled.} \label{fig:condweak_compare_limits} \end{figure}

In contrast to the case of potential scattering and the classical spin
approximation, which can be solved exactly,
the full quantum mechanical spin is investigated
by T-matrix resummed perturbation theory within
leading order in the cotunneling amplitude.
The numerically determined results for the nonlinear conductance
are summarized in Fig.~\ref{fig:condweak_compare_limits} for three
different regimes of coupling asymmetry. For stronger coupling to
the S-lead, resonances similar to the Yu-Shiba-Rusinov states below the gap are very sharp and
the dip at the gap edge well defined. As the coupling to the
normal lead increases, these resonances become broader and start filling-in the dip.
Eventually the sub-gap resonances merge with the dip and reproduce the
usual BCS-profile at the gap edge.

\section{Model}

\subsection{Model system}
\label{sec:model}

We consider a quantum dot coupled to a normal and a
superconducting lead (cf.~Fig.~\ref{fig:Shiba_Andreev}a). The highest partially occupied orbital on
the dot is represented by a single-orbital Anderson model:
\begin{align}
H=H_{S}+H_{N}+H_{T}+H_{Dot},
\end{align}
with
\begin{align}
H_{\alpha}=&\sum_{\mathbf{k}\sigma}(\varepsilon_{k}-\mu_{\alpha})
c^{\dagger}_{\alpha\mathbf{k}\sigma}
c_{\alpha\mathbf{k}\sigma}\notag\\
&+\sum_{\mathbf{k}}( \Delta_{\alpha}^{\ast}
c_{\alpha\mathbf{k}\uparrow}c_{\alpha-\mathbf{k}\downarrow}
+\Delta_{\alpha}
c_{\alpha -\mathbf{k}\downarrow}^{\dagger}
c_{\alpha \mathbf{k}\uparrow}^{\dagger})\notag\; ,\\
H_{T}=&\sum_{\alpha\mathbf{k}\sigma}\left(t_{\alpha}
c_{\alpha\mathbf{k}\sigma}^{\dagger}d_{\sigma}+
t_{\alpha}^{\ast}d_{\sigma}^{\dagger}
c_{\alpha\mathbf{k}\sigma}\right),\notag\\
H_{Dot}=&\sum_{\sigma}\varepsilon_{d}d_{\sigma}^{\dagger}d_{\sigma}+
U n_{d\uparrow}n_{d\downarrow},
\end{align}
where $\alpha=N, S$ labels respectively a normal metal electrode ($\Delta_N=0$)
and a
superconducting lead with an ordinary s-wave BCS DOS and a gap
 which is assumed to be real ($\Delta_S=\Delta$). This can be safely assumed, since the
phase of the superconductor does not play a role in an S/N junction.
Dot electrons of spin $\sigma$ are created by
$d^{\dagger}_{\sigma}$ in an orbital of energy $\varepsilon_{d}$ and
with a mutual Coulomb interaction strength $U$. The tunnelling
amplitude between dot and electrodes is denoted by $t_{\alpha}$, and
the Coulomb blockaded dot is tuned by the gate-voltage
($V_{g}\sim\varepsilon_{d}$) to hold a well-defined number of
electrons. For a partial filling of this highest lying dot-orbital
of one, i.e.~a single electron on the dot, we thus assume that
$\Gamma_{\alpha}=\pi\nu_{F}|t_{\alpha}|^{2}\ll
\max(-\varepsilon_{d},\varepsilon_{d}+U)$.

In order to represent the exchange interaction between conduction
electrons and the spinful quantum dot, it is necessary to augment
the standard BCS Nambu-spinors to liberate the spin, from the
charge. To this end, we introduce the four-spinors
\begin{eqnarray}
\psi_{\alpha \mathbf{k}}^{\dagger}&=&\left(
\begin{array}
[c]{cccc}
c_{\alpha\mathbf{k}\uparrow}^{\dagger}, &
c_{\alpha-\mathbf{k}\uparrow}, &
c_{\alpha\mathbf{k}\downarrow}^{\dagger}, &
c_{\alpha-\mathbf{k}\downarrow}
\end{array}
\right),\\
\psi_{\alpha \mathbf{k}}&=&\left(
\begin{array}
[c]{l}
c_{\alpha \mathbf{k}\uparrow}\\
c_{\alpha -\mathbf{k}\uparrow}^{\dagger}\\
c_{\alpha \mathbf{k}\downarrow}\\
c_{\alpha -\mathbf{k}\downarrow}^{\dagger}
\end{array}
\right),
\end{eqnarray}
satisfying the anti-commutation relations
\begin{eqnarray}
\{\psi_{\alpha \mathbf{k}\eta},
\psi_{\alpha '\mathbf{k}'\eta'}^{\dagger}\}&=&
\delta_{\alpha\alpha'}\delta_{\mathbf{k}\mathbf{k}'}
m_{\eta\eta^{\prime}}^{0},\\
\{\psi_{\alpha \mathbf{k}\eta}^{(\dagger)} ,\psi_{\alpha
\mathbf{k}'\eta'}^{(\dagger)}\}&=&
\delta_{\alpha\alpha'}\delta_{-\mathbf{k}\mathbf{k}'}
m_{\eta\eta'}^{c},\label{eq:anticommut}
\end{eqnarray}
where we have introduced the following set of $4\times4$ matrices:
\begin{align}
m^{0}&
=\left(
\begin{array}
[c]{cccc}
1 & 0 & 0 & 0\\
0 & 1 & 0 & 0\\
0 & 0 & 1 & 0\\
0 & 0 & 0 & 1
\end{array}
\right)  , \qquad
m^{a}
=\left(
\begin{array}
[c]{cccc}
1 & 0 & 0 & 0\\
0 & -1 & 0 & 0\\
0 & 0 & 1 & 0\\
0 & 0 & 0 & -1
\end{array}
\right), \notag\\
m^{b}&
=\left(
\begin{array}
[c]{cccc}
0 & 0 & 0 & -1\\
0 & 0 & 1 & 0\\
0 & 1 & 0 & 0\\
-1 & 0 & 0 & 0
\end{array}
\right), \qquad
m^{c}
=\left(
\begin{array}
[c]{cccc}
0 & 1 & 0 & 0\\
1 & 0 & 0 & 0\\
0 & 0 & 0 & 1\\
0 & 0 & 1 & 0
\end{array}
\right).
\end{align}
Within this notation, the Hamiltonian for the leads reads
\begin{align}
\label{eq:1}
H_{\alpha}=
\frac{1}{2}\sum_{\mathbf{k}}\psi_{\mathbf{k}}^{\dagger}\
\xi_{\mathbf{k}}m^{a}\ \psi_{\mathbf{k}}
+
\frac{1}{2}\sum_{\mathbf{k}}\psi_{\mathbf{k}}^{\dagger}\
\Delta_{\alpha}m^{b}\ \psi_{\mathbf{k}}\; ,
\end{align}
where $\alpha = S, N$, using $\Delta_N = 0$ and
$\Delta_S=\Delta$.

Since we focus entirely on cotunneling, we project out
charge-fluctuations by means of a Schrieffer-Wolff
transformation.~\cite{Schrieffer66,Salomaa88} This simplifies the
Anderson model to the effective cotunneling model:
\begin{align}
H_{cotun}=H_{exch}+H_{pot},
\end{align}
with
\begin{align}
H_{exch}= \frac{1}{4} \sum_{\alpha^{\prime}\mathbf{k}^{\prime}\eta',\alpha
\mathbf{k}\eta i} \  \sum_{i=x,y,z} \ J_{\alpha^{\prime}\alpha} S^{i}\psi_{\alpha^{\prime}\mathbf{k}^{\prime}\eta'}^{\dagger
}m_{\eta'\eta}^{i} \psi_{\alpha \mathbf{k}\eta},\label{eq:Kondo}
\end{align}
where the 4-spinor notation has been supplemented by the following augmentation of the Pauli matrices:
\begin{align}
m^{x}  &=\left(
\begin{array}
[c]{cccc}
0 & 0 & 1 & 0\\
0 & 0 & 0 & -1\\
1 & 0 & 0 & 0\\
0 & -1 & 0 & 0
\end{array}
\right),
\qquad
m^{y}=\left(
\begin{array}
[c]{cccc}
0 & 0 & -i & 0\\
0 & 0 & 0 & -i\\
i & 0 & 0 & 0\\
0 & i & 0 & 0
\end{array}
\right)  ,\nn \\  & \qquad  \qquad
m^{z}  =\left(
\begin{array}
[c]{cccc}
1 & 0 & 0 & 0\\
0 & -1 & 0 & 0\\
0 & 0 & -1 & 0\\
0 & 0 & 0 & 1
\end{array}
\right)  \;.
\end{align}
The spin exchange interaction, Eq.~\eqref{eq:Kondo},
applies only to a spinful quantum dot holding a net
spin,
$\vec{S}=\frac{1}{2}d^{\dagger}_{\sigma'}\vec{\tau}_{\sigma'\sigma}
d_{\sigma}$, whereas the potential scattering term
\begin{align}
H_{pot}= \frac{1}{2} \sum_{\alpha^{\prime}\mathbf{k}^{\prime}\eta',\alpha
\mathbf{k}\eta}W_{\alpha'\alpha} \psi_{\alpha'\mathbf{k}'\eta'}^{\dagger
}m_{\eta'\eta}^{a}\psi_{\alpha \mathbf{k}\eta} \; ,\label{eq:potscat}
\end{align}
applies to spinless quantum dots, as well as to spinful dots away from the particle-hole symmetric point in the middle of the relevant Coulomb blockade diamond. For a dot holding one electron, the two different cotunneling amplitudes are given by
\begin{align}
J_{\alpha\alpha'} &=\frac{2 U t_{\alpha} t_{\alpha'}^*}{(\varepsilon_{d}+U)(-\varepsilon_{d})},
\label{eq:J}
\end{align}
and
\begin{align}
W_{\alpha\alpha'} &= \frac{(2\varepsilon_{d}+U) t_{\alpha}
  t_{\alpha'}^*}{2(\varepsilon_{d}+U)(-\varepsilon_{d})}\; ,
\label{eq:W}
\end{align}
where indeed $W_{\alpha\alpha'}=0$ at the particle-hole
symmetric point $\varepsilon_{d}=-U/2$. This result is readily
generalized to any odd number of electrons on the dot, as long
as only single-electron charge fluctuations are being
eliminated by the Schrieffer-Wolff transformation. In the case of an even occupied dot, we denote the corresponding 
2nd order cotunneling amplitude by $W_{\alpha\alpha'}^{e}$. 

As an illustrative intermediate step, we shall also discuss the
case of a classical spin, where we replace the full exchange
cotunneling term by a spin-dependent potential scattering term:
\begin{align}
H_{exch}^{cl.}=\frac{1}{2} \sum_{\alpha'\mathbf{k}^{\prime}\eta',\alpha
\mathbf{k}\eta}W_{\alpha'\alpha}^{s} \psi_{\alpha'\mathbf{k}\eta'}^{\dagger
}m_{\eta'\eta}^{z} \psi_{\alpha \mathbf{k}\eta} \;.\label{eq:HexchCL}
\end{align}
We stress that this is merely a simplification of $H_{exch}$
corresponding to the limit of $J\to 0$ and $S\to\infty$,
keeping the product, $JS$, constant. This was in fact the
problem considered e.g~by Shiba~\cite{Shiba68} and as we shall
demonstrate it already captures the essential physics of the
spin-induced sub-gap resonances even for the full quantum spin.

\subsection{Unperturbed Green functions}

Since we want to study transport beyond the linear regime, we
do the perturbation theory using Keldysh formalism. We define
the contour-ordered conduction electron Green functions:
\begin{align}
G_{\alpha\mathbf{k}\eta,\alpha^{\prime}\mathbf{k}^{\prime}\eta^{\prime}}(\tau,\tau^{\prime})  &  =-i\left\langle
T_{C}\left\{  \psi_{\alpha\mathbf{k}\eta}(\tau)\psi_{\alpha^{\prime}\mathbf{k}^{\prime}\eta^{\prime}}^{\dagger}
(\tau^{\prime})\right\}\right\rangle.
\end{align}
In all expressions
encountered below, the Green functions are summed over
momentum. Therefore we start out by stating the unperturbed
momentum summed Green functions for the leads. We have for the
retarded and advanced Green functions
\begin{align}
G^{(0),R/A}(\omega) &  = \sum_{\mathbf{k}}G^{(0),R/A}(\mathbf{k},\omega)
\nn \\
&= - \pi\nu_{F}\theta(D-|\omega|)
\frac{(\omega \pm i 0^+) m^{0}+\Delta m^{b}}{\sqrt{\Delta^{2} - (\omega \pm i  0^+)^{2}}},
\end{align}
where $D$ denotes the conduction electron half-bandwidth and we assume a constant DOS (pr.~spin), $\nu_{F}$. For the spectral function, we then have
\begin{align}
A^{(0)}(\omega) & = 2\pi\nu_{F}\theta(D-|\omega|)\theta(|\omega|-|\Delta|)
\text{sign}(\omega) \nn \\ & \quad \times
\frac{\omega m^{0}+\Delta m^{b}}{\sqrt{\omega^{2}-\Delta^{2}}}.
\end{align}
Notice that the anomalous part of the DOS is an odd
function of $\omega$, which falls off as $\Delta/\omega$ for large $\omega$. For the normal lead, where $\Delta=0$, the spectral function reduces to
\begin{align}
A_N^{(0)}(\omega) &= 2 \pi \nu_F \theta(D - |\omega|)\; .
\end{align}
From the momentum-integrated spectral function, the lesser
and greater Green functions are readily obtained as
\begin{align}
G^{(0),<}(\omega) &  =iA(\omega)f(\omega),
\end{align}
and
\begin{align}
G^{(0),>}(\omega) &  =-iA(\omega)(1-f(\omega)),
\end{align}
where $f$ denotes the Fermi function. The voltage is applied to
the normal lead only, thus avoiding the complication with a
running phase in the superconducting lead, and therefore the
voltage $V$ enters only in the normal lead Green functions
\begin{align}
  G_{N,\eta\eta'}^{(0),<} &= i f(\omega - m^a_{\eta\eta} V) m^0_{\eta\eta'} A_N^{(0)}(\omega), \\
  G_{N,\eta\eta'}^{(0),>} &= - i (1 - f(\omega - m^a_{\eta\eta} V)) m^0_{\eta\eta'} A_N^{(0)}(\omega),
\end{align}
where the chemical potential shift has opposite sign for particle and hole components.

\subsection{Current and T-matrix}

Since we will be particularly interested in bound states (or
resonances) as poles in the conduction electron T-matrix, it is
convenient to express the current in terms of the T-matrix. To
do this, we momentarily revert to the underlying Anderson model
for which the current operator is found as the rate of change
of the number of particles in the superconducting
lead:
\begin{align} \hat{I}  & =\partial_{t}(Q_{S})
=\frac{(-e)}{2i\hbar}\sum_{\mathbf{k}}[H_T,\widehat
\psi_{S\mathbf{k}}^{\dagger}m^{a} \widehat\psi_{S\mathbf{k}}].
\end{align}
Introducing four-spinors for the dot electrons:\begin{equation}
\phi^{\dagger}=\left(
\begin{array}
[c]{cccc}
d_{\uparrow}^{\dagger}, & d_{\uparrow}, & d_{\downarrow}^{\dagger}, &
d_{\downarrow}
\end{array}
\right)  ,\;\;\;\;\;\;\;\phi=\left(
\begin{array}
[c]{l}
d_{\uparrow}\\
d_{\uparrow}^{\dagger}\\
d_{\downarrow}\\
d_{\downarrow}^{\dagger}
\end{array}
\right),
\end{equation}
the tunneling term takes the following form:
\begin{align}
H_{T}=\frac{1}{2}\sum_{\alpha,\mathbf{k},\eta} t_{\alpha}m_{\eta\eta}^{a}\Big(
\hat \psi_{\alpha \mathbf{k}\eta}^{\dagger} \phi_{\eta}
+\phi_{\eta}^\dagger \hat \psi_{\alpha \mathbf{k}\eta}
\Big),
\end{align}
where the tunneling amplitude has been chosen to be real, which
is always possible for a single-level model with one normal
lead, since a phase can be absorbed by a gauge transformation.

The expectation value of the current operator now involves the
mixed $4\times 4$ Nambu Green functions, $i\langle
T_C\psi_{S,\eta'}(\tau)\phi^\dag_{\eta}(\tau')\rangle$, and
one can show that
\begin{align}
\langle\hat{I}\rangle
&= \frac{1}{2} \frac{e}{\hbar}\sum_{\gamma\eta}t_{S}^2 m_{\gamma\gamma}^a\int\frac{d\omega}{2\pi} \nonumber\\
&\times
\left(\big[G_{d;\eta\gamma}(\omega)      G^{(0)}_{S;\gamma\eta}(\omega) \big]^<
- \big[G^{(0)}_{S;\eta\gamma}(\omega)
G_{d;\gamma\eta}(\omega)\big]^<\right),
\label{eq:current_GdG}
\end{align}
where $G_{d;\eta\gamma}$ is the dot electron Green function in spinor space. In
Eq.~\eqref{eq:current_GdG} we use the shorthand $[A(\omega)B(\omega)]^{<}=
A^{R}(\omega)B^{<}(\omega)+A^{<}(\omega)B^{A}(\omega)$ implied
by Langreth rules and subsequent Fourier-transformation. From
equations of motion, the dot electron Green function can be
obtained from the conduction electron T-matrix as
\begin{align}
\label{eq:GD_Tmat}
t_{\alpha}^2
G^{R}_{d;\eta\eta'}(t, t')
&= m_{\eta\eta}^a m_{\eta'\eta'}^a
T^R_{\alpha;\eta\eta'}(t,t'),
\end{align}
which may be inserted to obtain the following formula for the current:
\begin{align}
\langle I \rangle&=\frac{e}{h} \sum_{\eta\gamma} m_{\eta\eta}^a
\int\limits_{-\infty}^\infty d\omega\;
{\mathrm Re}\Big\{[T_{S;\eta\gamma}(\omega)
     G^{(0)}_{S;\gamma\eta}(\omega)]^<
\Big\},
\label{eq:current}
\end{align}
where we have used the relation ${\cal G}^<_{\eta\eta'} = -
({\cal G}_{\eta'\eta}^<)^\dag$. The current now relies solely
on the conduction electron T-matrix, which we calculate within
the effective cotunneling model derived from the
Schrieffer-Wolff transformation.
Thus the details of the quantum dot enter only via the
  interaction with the leads and in the following we study
  Eq.~\eqref{eq:current} for $H_{pot}$ and/or $H_{exch}$ as
defined in section~\ref{sec:model}.

The T-matrix effectively sums up an infinite repetition of the
one-particle irreducible self energy $\Sigma$, which we obtain either
exactly (for spinless dots and for the limit of a classical
spin) or to leading order perturbation theory in the exchange
cotunneling amplitude. In general, the retarded or advanced
$SS$ component of the conduction electron T-matrix is found as
\begin{align}
T_{SS}^{R/A}(\omega)
&= \Sigma^{R/A}_{SS, \eff}(\omega) \big(m^{0}-G^{(0),R/A}_S(\omega)
\Sigma^{R/A}_{SS, \eff}(\omega)\big)^{-1},
\label{eq:Tret}
\end{align}
with an effective $SS$ self energy which incorporates as well all processes going via the normal lead back into the superconducting lead:
\begin{align}
&\Sigma^{R/A}_{SS,\eff}(\omega) = \Sigma^{R/A}_{SS}(\omega) \notag
\\ & \quad + \Sigma^{R/A}_{SN}(\omega) \big[ [G_N^{(0),
  R/A}(\omega)]^{-1}-\Sigma^{R/A}_{NN}(\omega)\big]^{-1}\Sigma^{R/A}_{NS}(\omega).
\label{eq:Sigma_SSeff}
\end{align}
This effective self energy also enters the Dyson equation for the interacting $SS$ Green function
\begin{align}
G_{SS}^{R/A}(\omega) &=
\Big\{ [G^{(0), R/A}_{S}(\omega)]^{-1} -
\Sigma^{R/A}_{SS, \eff}(\omega) \Big\}^{-1}.
\end{align}

In the current formula Eq.~(\ref{eq:current}), the lesser
T-matrix enters through the following combination of Nambu
matrices (omitting the $\omega$-dependence in the following):
\begin{align}
&T_{SS}^< G_{S}^{(0),A} + T^R_{SS} G_{S}^{(0), <}
=
T^R_{SS} G_S^{(0), <}
+ \Sigma^<_{SS, \eff} G^A_{SS}
\nn \\ & \quad
+ T_{SS}^R \Big( G_{S}^{(0),<} \Sigma_{SS,\eff}^A
+
G_{S}^{(0),R} \Sigma_{SS,\eff}^<  \Big) G_{SS}^A
\nn \\ & \quad
+ T_{SN}^R \Big( G_{N}^{(0),<} \Sigma_{NS, \eff}^A
+
G_{N}^{(0),R} \Sigma_{NS,\eff}^<  \Big) G_{SS}^A,
\label{eq:tmatSS}
\end{align}
where
\begin{align}
\Sigma^{A}_{NS,\eff} &= \Sigma^{A}_{NS} + \Sigma^{A}_{NN} (m^0 -
 G^{(0), A}_N \Sigma^{A}_{NN})^{-1}   G^{(0), A}_N \Sigma^{A}_{NS},
\label{eq:SigmaA}\\
 \Sigma^<_{\alpha\alpha',\eff} &= \Sigma_{\alpha\alpha'}^< +
 \Sigma_{\alpha N}^< (m^0 - G_N^{(0),A} \Sigma_{NN}^A)^{-1} G_N^{(0),A} \Sigma_{N\alpha'}^A,
\label{eq:SigmaL}
\end{align}
and
\begin{align}
T_{SN}^R &=
\Big( m^0
+ T^R_{SS} G_{S}^{(0),R}
\Big) \Sigma_{SN}^R
\Big(m^0 - G_{N}^{(0),R} \Sigma_{NN}^R \Big) ^{-1}.
\label{eq:tmatSN}
\end{align}
The irreducible self energy $\Sigma_{\alpha\alpha'}$ depends on
the details of the interacting system, thus determining the
expressions for the T-matrix and the conductance as will be discussed
now for the various cases.

\section{Results - Exactly solvable}

\subsection{Spinless dot - potential scattering}
\label{sec:PS}

For a spinless dot, as for example one of even occupation with
a singlet ground state, the effective cotunneling model has
only the potential scattering term (\ref{eq:potscat}), and the
irreducible conduction electron self energy is then exact
already to first order in the potential scattering amplitude,
$W^{e}_{\alpha\alpha'}$:
\begin{align}
\Sigma_{\alpha\eta,\alpha'\eta'}^{e}
&=\frac{1}{2} m^a_{\eta\eta'}W^{e}_{\alpha\alpha'}\; .
\end{align}
Since the self energy is independent of frequency, it has
$\Sigma^R = \Sigma^A \equiv \Sigma^{e}$ and $\Sigma^{<}= 0$,
and the effective $SS$ Nambu matrix self energy therefore takes
the following simple form:
\begin{align}
\Sigma^{e,R/A}_{SS,\eff}
&=
\frac{1}{\pi\nu_{F}}(\sigma_{r}^{e}\ m^a\mp i \sigma_{t}^{e}\ m^0),
 \label{eq:Sigma_SSeff_PS}
\end{align}
where $\mp$ refers to retarded and advanced components and the
real dimensionless coefficients are defined by
\begin{align}
\sigma_t^{e} &=
\frac{\sigma^{e}_{SN} \sigma^{e}_{NS}
}{1 + (\sigma^{e}_{NN})^2},
\label{eq:sigma_t}\\
\sigma_{r}^{e}
&=\sigma^{e}_{SS} - \sigma^{e}_{NN}\ \sigma_t^{e},
\label{eq:sigma_r}
\end{align}
with the definition
\begin{align}
\sigma_{\alpha\alpha'}^{e}\equiv\frac{1}{2}\pi \nu_F W_{\alpha\alpha'}^{e}.
\end{align}
Here $\sigma_t$ is responsible for \textit{transmission} and
$\sigma_r$ can be related to \textit{reflections} back to the
superconducting lead. Note that all expressions are in terms of the
T-matrix $T_{SS}$ only. The T-matrix
$T_{NN}$ cannot be written in the simple fashion of
Eq.~\eqref{eq:Sigma_SSeff_PS} since superconducting correlations are
induced in the normal lead, whereas in $T_{SS}$ the coupling to normal
lead only modulates the already present anomalous contributions in the superconducting lead.

If the normal lead was decoupled from the quantum dot, i.e.~$W^{e}_{NS}=W^{e}_{SN}=0$,
we would have no transport, i.e.~$\sigma^{e}_t = 0$, and
$\sigma_{r}^{e}=\sigma_{SS}$. In the opposite limit where there
are no \textit{reflections}, i.e.~$W^{e}_{SS}=W^{e}_{NN} = 0$,
the system becomes equivalent to a tunnel junction, with
$\sigma_{r}^{e}=0$ and $\sigma_t^{e}=\sigma_{NS}\sigma_{SN}$.
In general, however, the cotunnel junction studied here
involves both numbers, $\sigma^{e}_{t/r}$. Even though the self
energy is exact in this case, it should be kept in mind that
the validity of our initial Schrieffer-Wolff transformation
relies on the fact that the four dimensionless numbers,
$\sigma_{\alpha\alpha'}$, are all much smaller than one.

Inserting the above self energy \eqref{eq:Sigma_SSeff_PS} into
Eqs.~(\ref{eq:tmatSS}-\ref{eq:tmatSN}) and using current
formula (\ref{eq:current}), we can now obtain a closed
analytical expression for the nonlinear conductance at zero
temperature. For voltages outside the gap, $V > \Delta $, we
have
\begin{align}
\frac{dI}{dV}&=
\frac{2e^2}{h}
\frac{4 \sigma_t^{e} }{\mathcal{D}^{e}_{V>\Delta}},
\label{eq:didvL}
\end{align}
with the denominator defined as:
\begin{align}
\mathcal{D}^{e}_{V>\Delta} &= 2 \sigma_t^{e} + (1 +
  (\sigma_t^{e})^2 + (\sigma_{r}^{e})^2)\ \sqrt{1 -( \Delta / V)^2}.
  \label{eq:Dg}
\end{align}
For voltages inside the gap, $V < \Delta$, we have instead
\begin{align}
\frac{dI}{dV}&=\frac{4 e^2}{h}
\frac{4 (\sigma_t^{e})^2}{\mathcal{D}^{e}_{V<\Delta}},
\end{align}
with denominator
\begin{align}
\mathcal{D}^{e}_{V<\Delta} &= (1 + (\sigma_t^{e})^2 +
  (\sigma_{r}^{e})^2)^2 ( 1 - (V/\Delta)^2 ) \notag\\
& \qquad + 4 (\sigma_t^{e})^2  (V/\Delta)^2.
\label{eq:Dl}
\end{align}

\begin{figure}[t]
\centering
\includegraphics[width=\columnwidth]{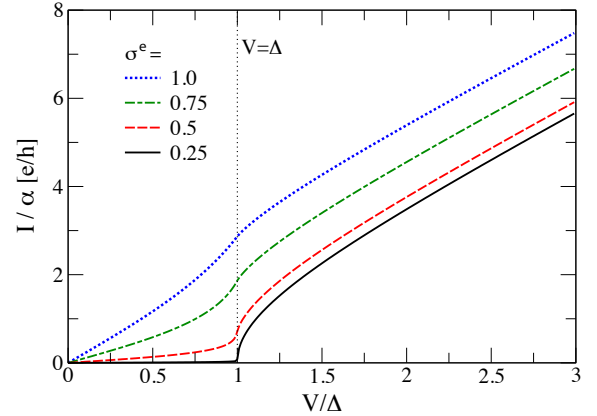}
\caption{Current $I/\alpha$ for the spinless dot (even occupation) with equal dimensionless
  cotunneling amplitudes to S and N given by respectively $\sigma^e
  \equiv \pi\nu_{F}W^{e}/2 = 1, 0.75, 0.5, 0.25$, corresponding to a
  cotunnel junction transmission of $\alpha = 0.8, 0.56, 0.2, 0.015$. Dotted line indicates the superconducting gap $V = \Delta$.}
 \label{fig:currentPS}
\end{figure}
In the limit of $\Delta\to 0$ or $V \to \infty$, the
differential conductance reaches the value $\frac{2e^2}{h}\alpha$, with a
\textit{cotunnel junction} transmission
\begin{align}
\alpha &=\frac{4\sigma_t^{e}}{(1+\sigma_t^{e})^2+
(\sigma_{r}^{e})^2}\;,
\label{eq:alpha}
\end{align}
which differs from that of a tunnel
junction~\cite{Cuevas96,Blonder82} merely by the presence of
the reflection terms $W^{e}_{SS}$ and $W^{e}_{NN}$ comprising
$\sigma_{r}^{e}$. Note that a rewriting of the full nonlinear
conductance, (\ref{eq:didvL}-\ref{eq:Dl}), in terms of this
$\alpha$ makes it identical to the expression found in
Ref.~\onlinecite{Cuevas96} for a tunnel junction, and to the
one by BTK~\cite{Blonder82}, when
expressing their barrier parameter $Z$ in terms of the
transmission, $Z^2 = \alpha^{-1} - 1$.

The conductance at $V=\Delta$ is exactly $4e^2/h$, whereas the zero-bias conductance is given by
\begin{align}
\frac{dI}{dV}\Big{|}_{V=0}&=\frac{4e^2}{h}
\left(
\frac{2\sigma_t^{e}}
{1+(\sigma_t^{e})^2+(\sigma_{r}^{e})^2}
\right)^2,
\end{align}
which agrees with the general result for the linear conductance of
an N/S interface~\cite{Beenakker92}: $G=4e^2/h\
\alpha^2/(2-\alpha)^2$, valid for any transmission $\alpha$.

\begin{figure}[t]
\centering
\includegraphics[width=\columnwidth]{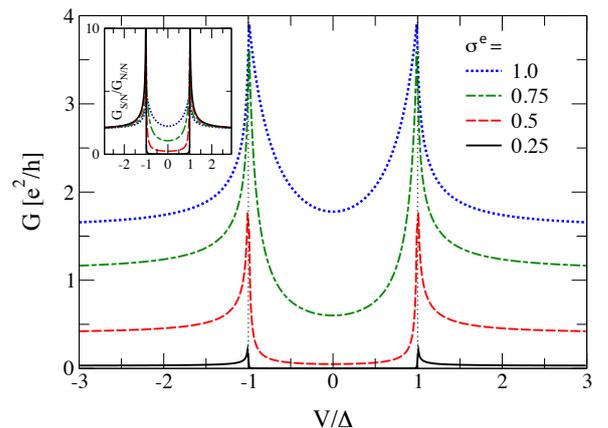}
\caption{Differential conductance  $G = dI/dV$ for
  the even dot with same parameters as in Fig.~\ref{fig:currentPS}.
  Inset: the nonlinear conductance $G_{S/N}$ normalized by the value of
  $G_{N/N} = 2 e^2/h\ \alpha$.}\label{fig:condPS}
\end{figure}

Sub-gap transport for $V<\Delta$ is allowed by Andreev
scattering processes, which proliferate with increasing
cotunneling amplitudes $W^{e}_{\alpha\alpha}$. This is clearly
seen in Fig.~\ref{fig:currentPS} where we plot the I-V curves
for different tunneling amplitudes. The corresponding nonlinear
conductances are shown in Fig.~\ref{fig:condPS} and seen to simply
reflect the BCS DOS for the smallest chosen
tunneling amplitude.

\subsection{Classical spin - spin dependent potential scattering}
\label{sec:classical_spin}

We commence with an extension of the spinless dot, described by a potential scattering term $W^{e}$,  to a dot holding a classical spin, which we describe by Eq.~\eqref{eq:HexchCL} in terms of a spin-dependent potential scattering term $W^{s}$. In this case, the exact irreducible self energy is given by
\begin{align}
\Sigma_{\alpha\eta,\alpha'\eta'}^{s}
&= \frac{1}{2}m^z_{\eta\eta'}W^{s}_{\alpha\alpha'}\;.
\end{align}
In line with the spinless case, the effective $SS$ self energy can be written as
\begin{align}
\Sigma^{s,R/A}_{SS,\eff}
&=\frac{1}{\pi\nu_{F}}
(\sigma_{r}^{s}\ m^z \mp i \sigma_{t}^{s}\ m^0)\; ,
\end{align}
where the same definitions \eqref{eq:sigma_t}  and \eqref{eq:sigma_r}
apply to the dimensionless
coefficients, $\sigma_{t/r}$, when simply replacing
$W^{e}_{\alpha\alpha'}$ by $W^{s}_{\alpha\alpha'}$. Note that
  the spin symmetry is broken by $m^z$ in contrast to the potential
  scattering case.

\begin{figure}[t]
\centering \includegraphics[width=\columnwidth]{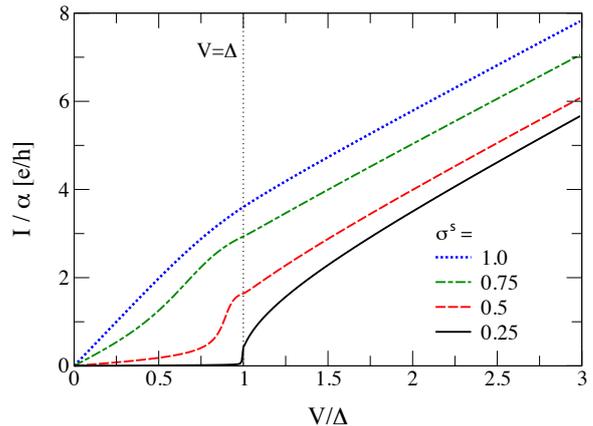}
\caption{Current $I/\alpha$ for a symmetrically coupled quantum dot occupied by
  an odd number of electron where the spin is treated classically. The
  coupling strengths are chosen $\sigma^s \equiv \pi\nu_{F}W^s/2= 1,
  0.75, 0.5, 0.25$ similarly to Fig.~\ref{fig:currentPS}.}\label{fig:currentCLspin}
\end{figure}

\begin{figure}[b]
\centering\includegraphics[width=\columnwidth]{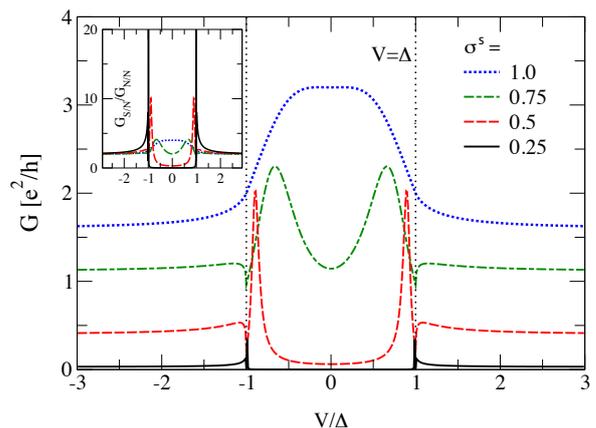}
\caption{Differential
  conductance $G = dI/dV$
  for the odd occupied dot with a classical spin where the coupling strengths are the same as in Fig.~\ref{fig:currentCLspin}. Inset: Conductance $G_{S/N}$ normalized by $G_{N/N} = 2e^2/h\ \alpha$.}\label{fig:condCLspin}
\end{figure}

As before, we can find a closed expression for the nonlinear
conductance at zero temperature. In
Figs.~\ref{fig:currentCLspin} and~\ref{fig:condCLspin} we first
show the resulting I-V curves and corresponding nonlinear
conductance for the same coupling strengths as used for the
spinless case in Figs.~\ref{fig:currentPS}-\ref{fig:condPS}. We
limit the plots to the case of symmetric couplings and return
to investigate asymmetric couplings for the full quantum
mechanical spin in the next section. For weak coupling, the
conductance in Fig.~\ref{fig:condCLspin} is similar to the
potential scattering case and merely reflects the BCS density
of states. For increasing coupling strength a sub-gap peak
appears symmetrically around zero bias and the conductance peak
at $V=\Delta$ changes to a dip, as discussed below. As the peak
moves closer to zero energy, it also becomes broader for our
choice of symmetrically coupled leads. Note that for very
strong coupling the two Yu-Shiba-Rusinov states mix and one sees only a
broad peak centered at zero voltage, which we should emphasize
has nothing to do with a Kondo resonance. All of these features
can be identified in the following analytical formulas for the
nonlinear conductance.

As in the case of potential scattering we find for the classical spin
exact expressions for the differential conductance from perturbation theory.
For voltages outside the gap, $V > \Delta $, we have
\begin{align}
  \frac{dI}{dV} &=   \frac{2 e^2}{h}\ 4\sigma_t^{s}\
\frac{\mathcal{D}_{V>\Delta}^{s}}
{4 (\sigma_{r}^{s})^2 (\Delta/V)^2
+ \big( \mathcal{D}_{V>\Delta}^{s}\big)^2},
\end{align}
with $\mathcal{D}_{V>\Delta}^{s}$ given by (\ref{eq:Dg}) with
$\sigma^{e}$ replaced by $\sigma^{s}$. In the limit $\Delta
  \to 0$ or $V \to \infty$ we find again that the conductance is given
by $2 e^2/h \ \alpha$ with the cotunnel junction transmission,
Eq.~\eqref{eq:alpha},  in terms
of $\sigma_{t/r}^s$. However, due to the term with $\sigma_r^s$ in the
denominator, the differential conductance here cannot be expressed in terms of $\alpha$ alone.

For $\sigma_{r}^{s} \approx 0$, i.e.~weak coupling to the
superconducting lead $\sigma_{SS} \approx 0$, the conductance outside
the gap is identical to that in the spinless case with a BCS-like
square root singularity at $|eV| \approx \Delta$. This changes rapidly
with increasing $\sigma_{r}^{s}$, which changes the singularity at
$|eV| = \Delta$ into a dip with a square root increase with voltage
(cf. Figs.~\ref{fig:currentCLspin} and \ref{fig:condCLspin}). At $V =
\Delta$ the differential conductance takes the value
\begin{align}
\frac{dI}{dV}\Big{|}_{V=\Delta}&=\frac{4e^2}{h}
\frac{(\sigma_{t}^{s})^{2}}{(\sigma_{r}^{s})^{2}+(\sigma_{t}^{s})^{2}},
\end{align}
which reaches the $4e^{2}/h$, attained in the spinless case, when
$\sigma_{r}^{s}\ll\sigma_{t}^{s}$, i.e.~in the case where the dot is
coupled much stronger to the normal, than to the superconducting
lead. On the contrary, for strong coupling to the superconducting lead,
$\sigma_r^s \gg \sigma_t^s$, the value of the conductance is
suppressed and the dip is clearly visible.

The spectral weight which has been removed at the gap edge can
instead be found inside the gap. For voltages inside the gap,
$V<\Delta$, the conductance takes the following form:
\begin{align}
 \frac{dI}{dV}
&= \frac{4 e^2}{h}
\Big(  2 V^2 - \Delta^2
+ \frac{\Delta^2}{4 (\sigma_{r}^{s})^2} \mathcal{D}_{V<\Delta}^{s} \Big)
\nn \\ & \qquad \times
\frac{\Gamma_S}{(V - \omega_{S})^2 + \Gamma_S^2}
\frac{\Gamma_S}{(V + \omega_{S})^2 + \Gamma_S^2},
\label{eq:condCLspin_inside}
\end{align}
 where
\begin{align}
  \omega_S &= \pm\Delta \frac{1 - ((\sigma_{r}^{s})^2 + (\sigma_t^{s})^2)^2}{(1 + (\sigma_{r}^{s})^2
  - (\sigma_t^{s})^2)^2 + 4 (\sigma_{r}^{s})^2 (\sigma_t^{s})^2}\; ,
\label{eq:CL_DeltaS} \\
  \Gamma_S &= \Delta
\frac{4 \sigma_{r}^{s} \sigma_t^{s}}{(1 + (\sigma_{r}^{s})^2
  - (\sigma_t^{s})^2)^2 + 4 (\sigma_{r}^{s})^2 (\sigma_t^{s})^2}\; .
\end{align}
The sub-gap conductance shows two peaks of approximately Lorentzian form, centered at energies $\pm\omega_S$ and having a width of $\Gamma_S$. In the case of vanishing coupling to the normal lead, i.e.~$\sigma_{t}^{s}=0$ and $\sigma_{r}^{s}=\sigma_{SS}^{s}$, the resonances sharpen to form real bound states, located at:
\begin{align}
\omega_{S}&=\pm\Delta\frac{1-(\sigma_{r}^{s})^2}{1 + (\sigma_{r}^{s})^2}.
\label{eq:Shiba_classical}
\end{align}
This limit reproduces the case of a classical spin embedded in
a bulk superconductor.\cite{Yu65,Soda67,Shiba68,Rusinov69}. For weak coupling
this bound state is thus offset from the
superconducting gap by roughly the interaction strength. For
strong coupling the dependence changes, but interactions of the
order of the band width will on the other hand conflict with
the confinement of electrons on the quantum dot. The classical
spin case is artificial in the sense that the spin symmetry is
broken although the spin exchange interaction does not break
this symmetry. In the next section it is shown that spin-induced sub-gap bound states, qualitatively similar to those found in this section, are
still present if the spin is treated quantum mechanically.

\section{Results - Quantum spin}

\subsection{Exchange cotunneling}

Whereas the calculation for the classical spin is exact, we have to rely on leading order perturbation theory when it comes to calculating the conduction electron self energy in the case of exchange cotunneling with a quantum mechanical spin. For zero magnetic field, the leading term in the conduction electron self energy is of second order in the exchange interaction:
\begin{align}
\Sigma_{\alpha\eta,\alpha'\eta'}(t-t')
=&\frac{1}{16}\sum_{\g,\g'}\sum_{ij}
\langle S^i(t) S^j(t') \rangle_{(0)}
m_{\eta\g}^i m_{\g'\eta'}^j
\notag\\ &
\times \sum_{\alpha''}
J_{\alpha\alpha''}^{\eta\g}
G_{\alpha'',\g\g'}^{(0)}(t-t')
J_{\alpha''\alpha'}^{\g'\eta'}\; .\label{eq:exchselfen}
\end{align}

Before presenting the general results, let us first take a look at the
limit of negligible coupling to the normal lead, i.e.~$J_{NS}\approx 0$. With the experience from last section, we expect to find bound states showing up as zeros in the denominator of the T-matrix. These are located at energies $\omega_{S}$ for which
\begin{align}
0 &= \mathrm{det} \Big\{ [G^{(0), R/A}_{S}(\omega_{S})]^{-1} -
\Sigma^{R/A}_{SS}(\omega_{S}) \Big\},
\label{eq:determine_Shiba}
\end{align}
and inserting the self energy (\ref{eq:exchselfen}), we can identify bound states at energies
\begin{align}
\omega_{S}&=\pm\Delta\frac{1-3/4\, g_{SS}^2}{1+3/4\, g_{SS}^2}\; .
\label{eq:Shiba_energy}
\end{align}
in terms of the dimensionless exchange cotunneling amplitude
\begin{align}
g_{SS}=\frac{1}{4}\pi\nu_F J_{SS}\; .
\end{align}
Expression~\eqref{eq:Shiba_energy} is also valid for the general case
of a spin exchange interaction~\eqref{eq:Kondo} with a higher spin
(e.g.~spin-$1$ in an even dot) by replacing $3/4$ with $S(S+1)$.

The bound state energies in Eq.~\eqref{eq:Shiba_energy} match those
found for the classical spin \eqref{eq:Shiba_classical}  (in the limit
of vanishing coupling to the normal lead) if one replaces $W^s_{SS}$
by $\sqrt{3}/4\,  J_{SS}$. However, this analogy to the classical (or polarized) spin case does not hold anymore when the coupling to the normal lead is non-zero. We shall return to this comparison in Section~\ref{QuantSpin}.

As for the classical spin, a finite coupling to the normal lead
broadens the bound states into resonances. Unlike for the classical
spin, however, we cannot give a closed analytical expression for the
T-matrix nor for the nonlinear conductance.
\begin{figure}[htb]
  \centering
  \includegraphics[width=\columnwidth]{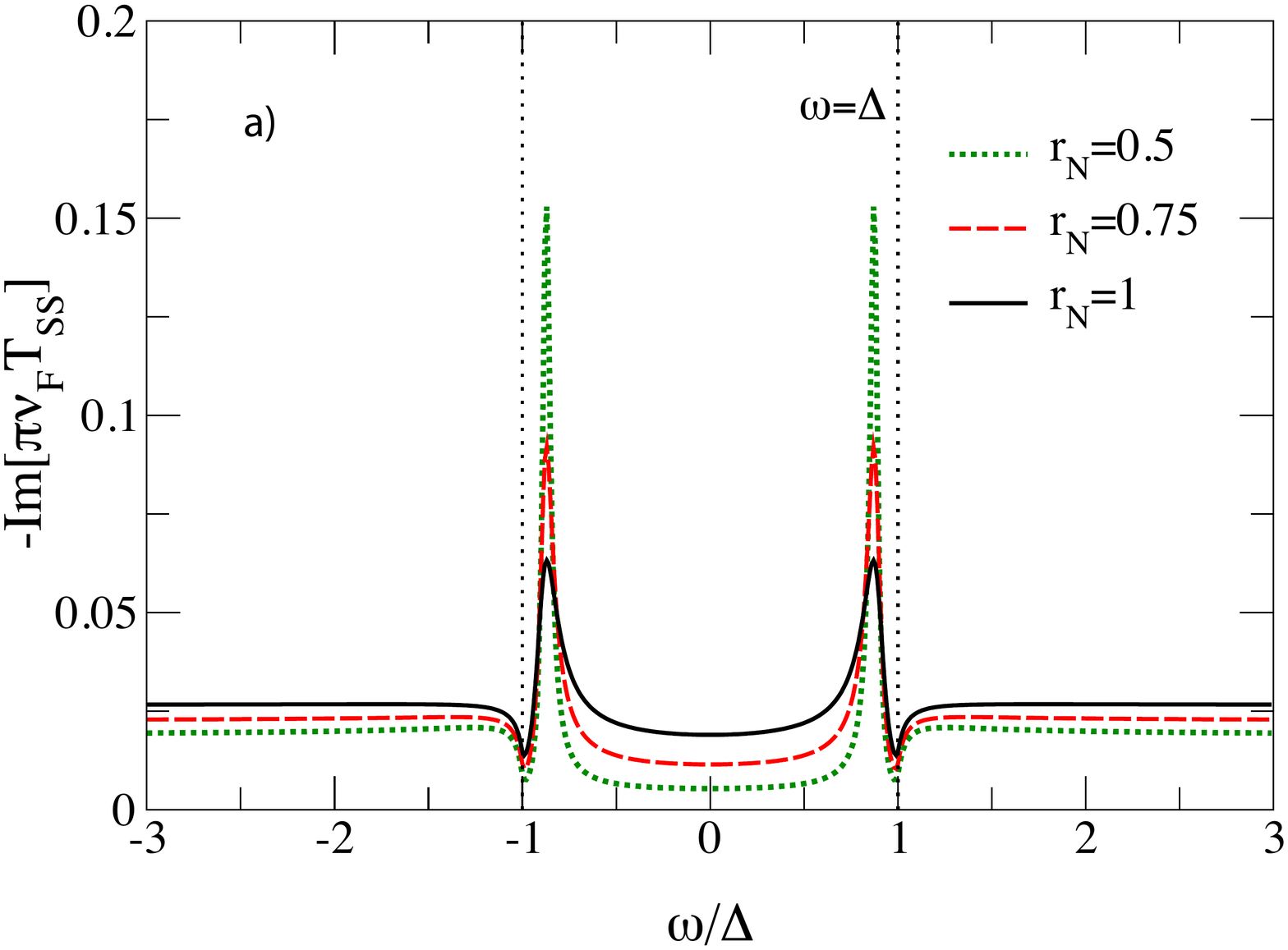}\\
  \includegraphics[width=\columnwidth]{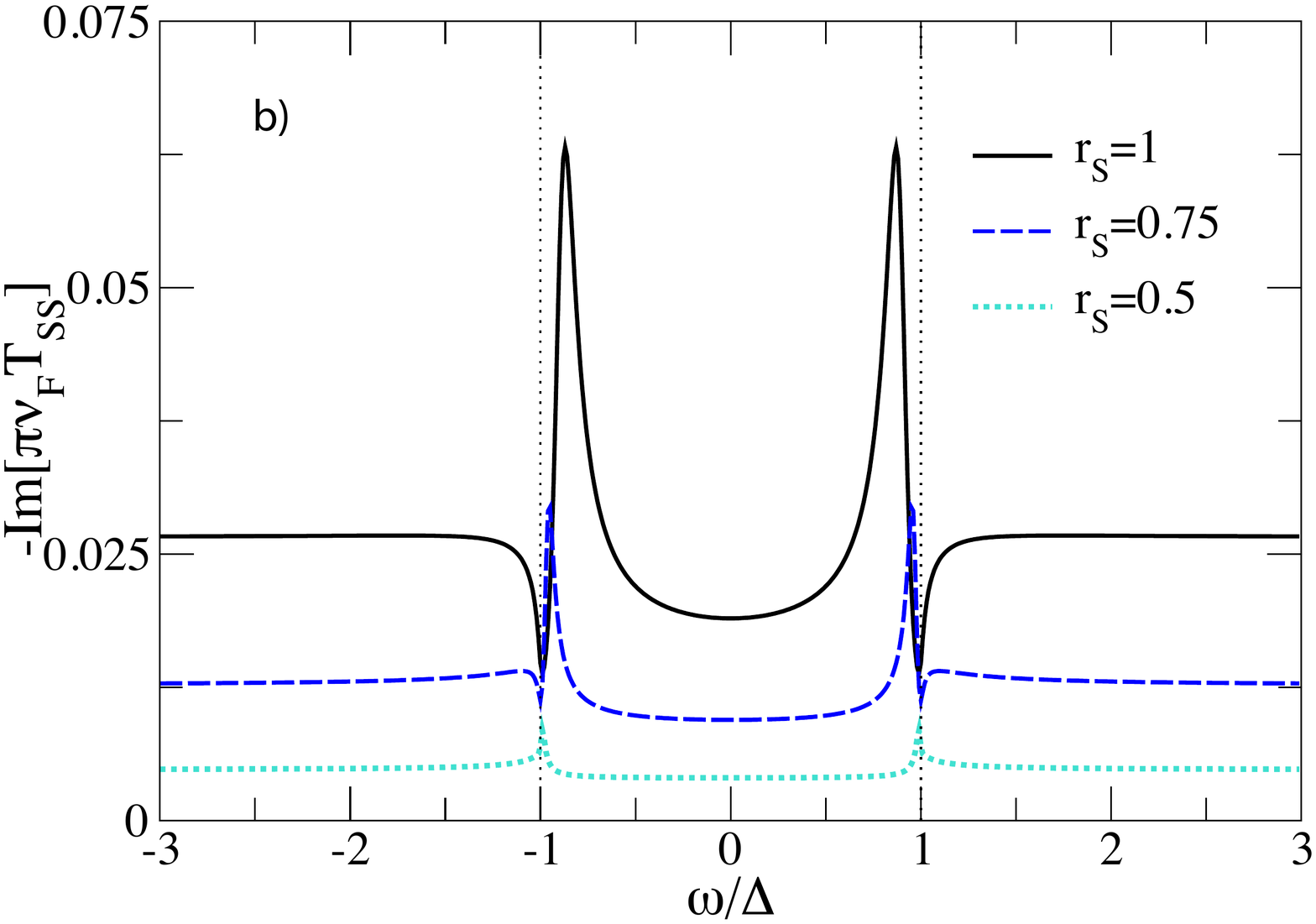}
  \caption{Imaginary part of the T-matrix, - Im$[\pi \nu_F T_{SS}]$,
    for a spinful quantum dot asymmetrically coupled
    to a superconducting and normal lead with
    $g_{0} \equiv \pi \nu_F J_0/4 = 0.3$
    leading to $\omega_{S} \approx 0.8735\Delta$ for
    $r_N = 0$;
    \textbf{a)} $r_S = 1$ is kept constant and the coupling to the
    normal lead $r_N$ is changed;
    \textbf{b)} $r_N = 1$ is kept constant and the coupling to the
    superconducting lead is reduced.
}
 \label{fig:Tmat}
\end{figure}
To obtain the retarded T-matrix, we insert \eqref{eq:exchselfen} into
\eqref{eq:Tret} and \eqref{eq:Sigma_SSeff}. The imaginary part of the
T-matrix, which is proportional to the local DOS on the
dot (cf. Eq.~(\ref{eq:GD_Tmat})), is plotted in Fig.~\ref{fig:Tmat}.
We parametrize the exchange cotunneling amplitudes from Eq. \eqref{eq:J} by
$J_{\alpha\alpha'}=J_0 r_{\alpha} r_{\alpha'}$, where $J_0 = 4 g_0 / \pi \nu_F$ and $r_\alpha =
t_\alpha/t_0$ is the ratio between the tunneling amplitude to lead
$\alpha = S,N$ and $t_0=\max\{t_N,t_S\}$.

The spin-induced sub-gap resonance is seen to stay at the same position as long as
the coupling to the superconducting lead is stronger than the coupling
to the normal lead, i.e.~$1=r_S \geq r_N$ in Fig.~\ref{fig:Tmat}a.
The peak though gets broader and
lower with increasing coupling, $r_N$, to the normal lead. In Fig.~\ref{fig:Tmat}b we show the behavior when the spin is stronger coupled to the normal lead by reducing the coupling to the superconducting lead $1 = r_N \geq r_S$. Besides a strong suppression of the overall value, we also observe that the resonance moves out towards the gap edge and, for very weak coupling to the superconducting leads, eventually gives back spectral weight to reconstruct the usual square root divergence at the superconducting gap $\Delta$.
These characteristics will be present again in the
nonlinear conductance through the quantum dot as discussed later on in this section.

\subsubsection{Gate dependence of the spin-induced sub-gap resonance energy}

At the particle-hole symmetric point, $\varepsilon_{d}=-U/2$, the spinful dot derived from the Anderson model has no potential scattering term, i.e.~$W=0$ at this point (cf. Eq.~\eqref{eq:W}). Leaving this point by adjusting the gate-voltage (which is proportional to $\varepsilon_{d}$), however, the potential scattering term has to be included, and it is clear from Eqs. (\ref{eq:J}-\ref{eq:W}) that both $J$ and $W$ will in fact increase in magnitude until eventually the Schrieffer-Wolff transformation breaks down as one comes too close to a charge-degeneracy point for the Coulomb blockaded quantum dot.

To investigate the dependence of the resonance frequency, $\omega_{S}$, on gate-voltage, we again consider the case where the dot is coupled only to the superconducting lead. As before, this is found as a root for the denominator in the T-matrix (cf. Eq.~\eqref{eq:determine_Shiba}), but now we have to include also the first order term from potential scattering, $\Sigma^{(1)}=
W_{\alpha\alpha'}m^{a}/2$, in the irreducible self energy. Doing this, one finds that the T-matrix pole will be located at
\begin{align}
  \label{eq:Shiba_plus_pot}
  \omega_{S} &= \pm \Delta \sqrt{\frac{(1 - 3/4\, g_{SS}^2)^2 +
      w_{SS}^2}{(1 + 3/4\, g_{SS}^2)^2 + w_{SS}^2}}\; ,
\end{align}
where $w_{SS}=\pi \nu_F W_{SS}/2$. For $w_{SS} = 0$ and $S=1/2$, we
thus reproduce Eq.~\eqref{eq:Shiba_energy} for a spin coupled to a
superconductor and for $g_{SS} = 0$ we find $\omega_{S} = \pm \Delta$,
i.e.~no spin-induced sub-gap state if there is no exchange coupling and only potential scattering as discussed in section~\ref{sec:PS}.

\begin{figure}[t]
\centering \includegraphics[width =\columnwidth]{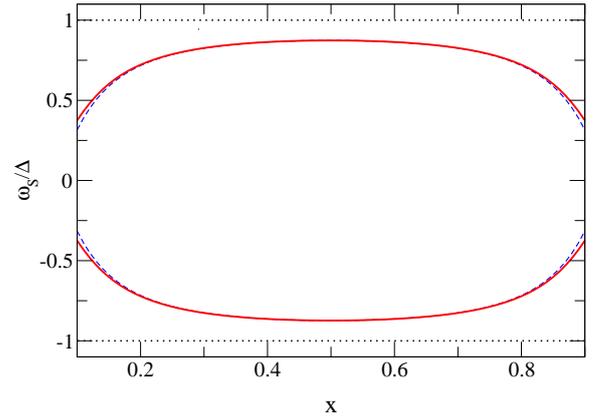}
\caption{Gate dependence of the spin-induced sub-gap resonance
  $\omega_S/\Delta$ in
  terms of $x=|\epsilon_d|/U$, and with $g_{SS} = 0.3$; red (full) line
  including potential scattering and blue (dashed) line without. Black
  (dotted) line indicates the energy of the superconducting gap.}
 \label{fig:gate_dep}
\end{figure}
In Fig.~\ref{fig:gate_dep} we show how the position of the sub-gap state
changes inside the Coulomb diamond as a function of $x =
|\epsilon_d|/U$.
As the coupling strength defined in Eq.~\eqref{eq:J} increases
towards the edges of the diamond (i.e.~$x=0$ and $x=1$), the bound
state moves closer to zero. At the same time we of course expect the
width of the peak to increase both due to the added influence of the
potential scattering term but first and foremost due to the increase
in cotunneling amplitudes as one moves away from the particle-hole
symmetric point ($x=1/2$). Notice that the potential scattering term,
with the gate dependence as defined in Eq.~\eqref{eq:W},
makes practically no difference until perturbation theory breaks down
anyway.
Interestingly, the gate dependence shown in Fig.~\ref{fig:gate_dep} is
very similar to that reported in recent experiments on
N/QD/S~\cite{Deacon10a} and S/QD/S junctions.~\cite{Grove09}

For the rest of the paper we solely deal with the particle-hole symmetric point where the potential scattering is zero.

\subsection{ Transport via spin-induced sub-gap resonances}
\label{QuantSpin}

To derive the nonlinear conductance in the case of a quantum mechanical spin, we need to include the full Keldysh matrix structure in Eq.~\eqref{eq:current} and \eqref{eq:tmatSS}. Therefore we cannot provide an analytic expression for this case, but the calculations are straightforward and do not need extensive numerical effort.

\begin{figure}[t]
\centering
\includegraphics[width=\columnwidth]{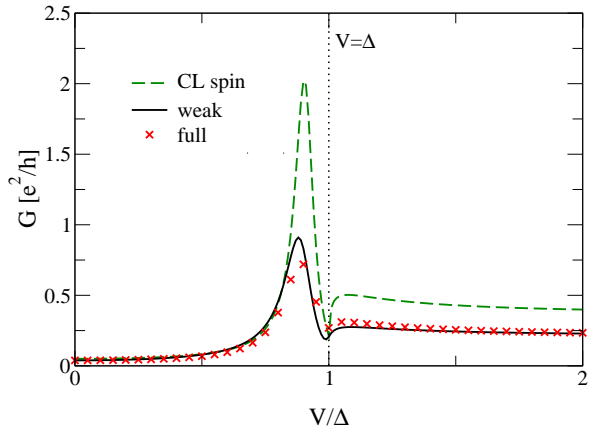}
\caption{Comparison
  of the differential conductance through a spinful quantum dot, treated
  quantum mechanically, taking into account all terms (red
  data points) or the the weak coupling term only (black line) for
  symmetric coupling $r_S/r_N = 1$ and $g_0 = 0.3$ . Furthermore, conductance
  treating the spin classically, dashed (green) line, with a coupling of $W^s_{\alpha\alpha'} = \sqrt{3}/4\ \sqrt{J_{\alpha\alpha'}}$.} \label{fig:cond_weak_full_CL}
\end{figure}

\begin{figure}[b]
\centering
\includegraphics[width=\columnwidth]{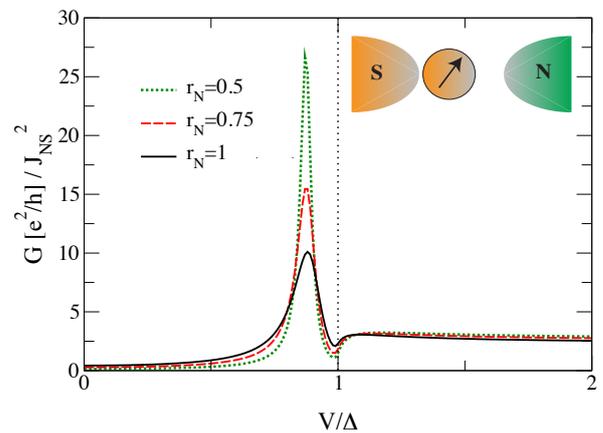}
\caption{Differential conductance $G = dI/dV$
  for increasing coupling to normal lead, $r_S = 1$ and $g_0 =
  0.3$. The conductance is renormalized by the ``transport'' coupling strength,
  $J_{NS}^2$, across the dot in order to
  compare the different curves. Inset: Illustration of a spin state
  coupled to the superconducting lead, thus forming a bound state,
  which is probed by the normal lead.}
\label{fig:condweak_change_rN}
\end{figure}

\begin{figure}[h]
\centering
\includegraphics[width=\columnwidth]{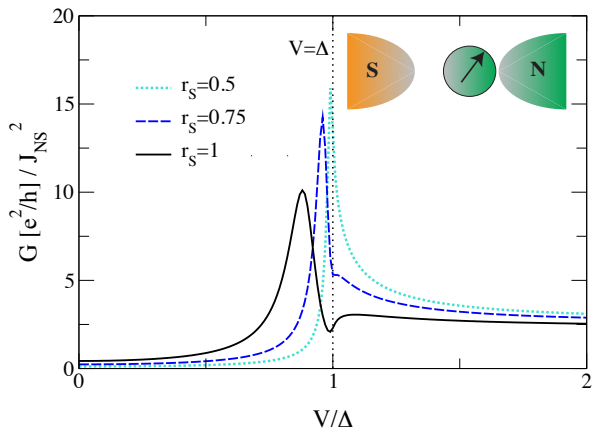}
\caption{
Renormalized nonlinear conductance $G/J_{NS}^2$ for increasing coupling to the superconducting lead and $r_{N}=1$; Inset: Illustration of a spin state strongly coupled to the
  normal lead and thus only the superconducting DOS is probed in transport.} \label{fig:condweak_change_rS}
\end{figure}

\begin{figure}[htb]
\centering
\includegraphics[width=\columnwidth]{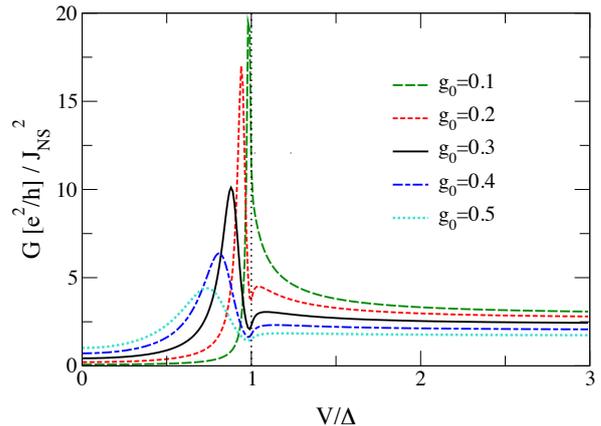}
\caption{Renormalized  differential conductance
  $G/J_{NS}^2$ for increasing coupling strength,  $g_0 = 0.1,
  0.2, 0.3, 0.4,$ and $0.5$, where both leads are symmetrically
  coupled, $r_N = r_S = 1$.} \label{fig:condweak_change_g0}
\end{figure}

\begin{figure}[tb]
\centering
\includegraphics[width=\columnwidth]{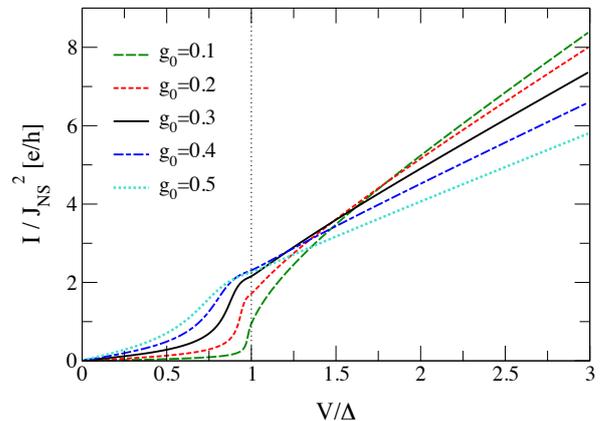}
\caption{Renormalized
  current $I/J_{NS}^2$  for the same parameters as
  Fig.~\ref{fig:condweak_change_g0}.}
 \label{fig:current_change_g0}
\end{figure}

We can distinguish between a hierarchy of contributions in
Eq.~\eqref{eq:tmatSS}. Summing over $m^a_{\eta\eta}$ in
Eq.~\eqref{eq:current}, $T^R_{SS} G^{(0),<}_S$ cancels out while
$\Sigma_{SS,eff}^{<} G_{SS}^A$ contributes to the current already to
second order in the spin exchange interaction $J_{\alpha\alpha'}$. This is the
dominant contribution in the conductance as illustrated in
Fig.~\ref{fig:cond_weak_full_CL}.
There ``weak coupling'' refers to
calculating the nonlinear conductance including $\Sigma_{SS, eff}^{<}
G^A_{SS}$ only, while ``full'' refers to including furthermore all contributions
in Eq.~\eqref{eq:tmatSS} which are of fourth order in lowest order in
$J_{\alpha\alpha'}$.
As can be seen in Fig.~\ref{fig:cond_weak_full_CL}, the weak coupling
term actually overestimates the height of the spin-induced sub-gap resonance and the
value at the edge of the superconducting gap slightly. However, for the coupling $g_0 =
0.3$ (cf. Fig.~\ref{fig:cond_weak_full_CL}) studied in the following,
this provides a very good approximation and
therefore Figs.~\ref{fig:condweak_change_rN}-\ref{fig:current_change_g0} are
calculated with the contribution from $\Sigma_{SS, eff}^{<} G_{SS}^A$ only. For stronger coupling the deviation is more severe. Furthermore, Kondo correlation effects have to be taken into account for strong coupling to the normal lead and this regime is not discussed here.

Fig.~\ref{fig:cond_weak_full_CL} also shows a comparison with
the conductance in the case of a spin treated classically. For
symmetrically coupled junctions we find that $W^s_{\alpha\alpha'} =
\sqrt{3}/4\ \sqrt{J_{\alpha\alpha'}}$ provides a reasonable agreement
although the energy of the  spin-induced sub-gap state is slightly shifted and the
value of the conductance is in general overestimated.
Whereas a linear relation, $W^s_{SS} = \sqrt{3}/4\ J_{SS}$,
is a good approximation for a spin
decoupled from the normal lead, a square root dependence
($W^s_{\alpha\alpha'} = \sqrt{3}/4 \sqrt{J_{\alpha\alpha'}}$)
fits the quantum mechanical case for symmetrically coupled junctions.
In practise, we can always find a value for the classical spin case,
which fits rather well the full spin-flip scattering case, but (as
already clear from the two simple limiting cases) there
is no obvious systematics involved and the value is strongly dependent
on asymmetry and strength of the coupling to the leads. Nevertheless,
we claim that the classical spin case provides good
qualitative insight into the problem of a spin coupled to a
superconductor.

Since the conductance is directly related to the local DOS, i.e.~Im$[G^A_{SS}]$, it is not
surprising, that sub-gap states are observed in the transport through
an N/QD/S cotunnel junction. As already illustrated in
Fig.~\ref{fig:Shiba_Andreev}, the local DOS of the
superconductor is probed by Andreev scattering processes to the normal
lead. The enhanced spectral density at the energy of the spin-induced sub-gap state leads to a
sub-gap peak in the differential conductance and a reduced DOS at
the superconducting gap $\Delta$ as reflected in $dI/dV$.
Comparing Figs.~\ref{fig:condweak_change_rN} and
\ref{fig:condweak_change_rS} with Fig.~\ref{fig:Tmat} we find
significant agreement, since Im$[T^R_{SS}]$ is (besides a prefactor
proportional to the interaction) given by the interacting local DOS.

As shown in Fig.~\ref{fig:condweak_change_rN} the spin-induced sub-gap state
stays at roughly the same energy $\omega_{S}$ given by Eq.~\eqref{eq:Shiba_energy}
(for a spinful quantum dot decoupled from the normal lead) as long as
the superconducting lead is stronger coupled to the dot than the
normal lead, $r_S \geq r_N$. This can be understood schematically as illustrated in the inset of Fig.~\ref{fig:condweak_change_rN}:
If the spin is coupled strongly to the superconducting lead, a
bound state is created and can be probed in transport through the N/S
setup with a weakly coupled normal lead.

On the contrary, if the normal lead is
stronger coupled to the impurity, $r_N \gg r_S$, we only probe the BCS superconducting
DOS (i.e. no spin-induced sub-gap resonances).
This is sketched
in the inset of Fig.~\ref{fig:condweak_change_rS}. Starting from a
 nonlinear conductance displaying a clear sub-gap peak for $r_N = r_S$,
the resonance moves closer towards the energy of the superconducting
gap $\Delta$ as the coupling to the normal lead is increased, $r_N > r_S$ in
Fig.~\ref{fig:condweak_change_rS}. If the
coupling to the normal lead dominates, no sub-gap resonance can be
distinguished from the square-root singularity at $\Delta$.
As was also illustrated in Fig.~\ref{fig:condweak_compare_limits} and is shown
again in Figs.~\ref{fig:condweak_change_rN} and
\ref{fig:condweak_change_rS} an experiment of the transport through an
N/QD/S junction can therefore have very different signatures depending on the
asymmetry of the coupling.

It is expected from the expression of $\omega_{S}$,
\eqref{eq:Shiba_energy}, that the spin-induced sub-gap bound state moves into the gap for
increasing coupling strength. This is illustrated in
Fig.~\ref{fig:condweak_change_g0} choosing  symmetric coupling to the
superconducting and normal lead.
For $g_0=0.1$ the sub-gap resonance state is present, but hidden at $V =
\Delta$. However, already for $g_0=0.2$ the conductance is seen to change into a peak
at $\omega_{S} \approx 0.94 \Delta$ and an associated dip at $\Delta$ instead of the square root divergence
of the clean superconducting DOS. Note that the conductance value at $\Delta$ decreases
with increasing coupling to the superconducting lead. Since we have chosen symmetric coupling in
Fig.~\ref{fig:condweak_change_g0}, the increasing coupling to the normal
lead causes a concomitant life time broadening of the peak inside the gap similar to the case of a classical spin.

Finally, Fig.~\ref{fig:current_change_g0} shows the current for the same
parameters as the conductance in Fig.~\ref{fig:condweak_change_g0}. The dip in the conductance at the
superconducting gap is hardly visible in the current, where it should
show up as a kink. Most interestingly, the current has a sharp
increase at the energy of the spin-induced sub-gap resonance $\omega_S$. This could easily
be misinterpreted as a reduced superconducting gap $\Delta$, since there is no
obvious difference between the line shapes of the curves with $g_0 = 0.1$ and $g_0
= 0.4$ in Fig.~\ref{fig:current_change_g0}. Therefore, a
clear distinction between a reduced superconducting gap and a
spin-induced sub-gap
resonance signature is best seen in the differential conductance.

\section{Summary and discussion}

We have investigated the transport characteristics of a Coulomb
blockaded quantum dot sandwiched between a normal and
a superconducting lead. The focus has been on the difference
between a dot with an even number of electrons and one with an
odd number of electrons or, in more general terms, the
difference between dots with zero or finite spin.

We have restricted the calculations to the cotunneling regime,
where charge fluctuations are strongly suppressed. The effective
model is in this situation a "cotunnel junction", where for
even occupancy (or a spinless dot) one obtains a simple
tunneling Hamiltonian for tunneling between N and S, plus
reflection terms (N to N, and S to S). In the language of the
Schrieffer-Wolff transformation, this is known as potential
scattering, see Eq.~\eqref{eq:potscat}. In contrast, for odd
occupancy (or a spinful dot) there is an additional term,
namely the Kondo, or exchange-cotunneling,
Hamiltonian, see Eq.~\eqref{eq:Kondo}, where
the tunneling electrons couple to the spin on the dot and hence
may induce spin flips. Furthermore, at the particle-hole
symmetric point, i.e.~in the middle of the odd occupancy
diamonds, the potential scattering term is absent and only the
Kondo Hamiltonian remains.

Because the effective even-occupied dot Hamiltonian is quadratic, the
current-voltage characteristic can be calculated exactly, and a
general expression that only depends on the transmission
(i.e.~the normal state conductance) has been derived here and also
previously in the literature for N/S
tunnel-junctions.~\cite{Blonder82,Beenakker92,Cuevas96}
For weak
transparency, the resulting differential conductance is
suppressed inside the gap  and develops the well-known
BCS square-root singular DOS at the gap edge.
With larger transparency the conductance is finite inside the
gap due to Andreev reflections, in full accordance with the BTK
model.\cite{Blonder82}

The problem with a spinful odd-occupied dot Hamiltonian cannot be solved analytically,
and in order to study the influence of the quantum spin
we calculate the lead electron self energy to second order in
perturbation theory which is subsequently summed in
the $T$-matrix. For this purpose, a general
expression for the current in terms of the lead electron
$T$-matrix has been derived. For weak tunnel couplings
(also weak enough that Kondo physics is not relevant, $T_K\ll
\Delta$), this approximation captures the important physics.
Interestingly, we show that the calculation gives results
which are qualitatively similar to a ``classical" approximation, where the
spin operator is replaced by a static magnetic moment. The
static spin approximation could for example result from a mean-field approach with an (unjustified) breaking of spin-rotational symmetry.
Nevertheless, the classical spin model provides good insight
by analogy to the well-known Yu-Shiba-Rusinov
bound states, which appear when electrons in a superconductor
scatter off a magnetic impurity. In the classical case, the bound
state energy can be determined exactly, and is located
inside the gap. For the quantum spin case there is
also a bound state inside the gap, but it does not rely on the unphysical
assumption of broken spin rotation symmetry. One may understand
the origin of this spin-induced sub-gap bound state from dynamically reduced superconducting correlations
due to the presence of an uncompensated magnetic spin. A Cooper pair consisting
of a spin-up and a spin-down electron is roughly speaking both repelled and attracted by
the impurity. Thus when breaking a Cooper pair, the energy of the localized excited
state is smaller than the superconducting gap by roughly the exchange
energy as a quasiparticle can gain energy by a spin-flip process.

The spin-induced resonance states have profound consequences for the
transport characteristics, because they give rise to a sub-gap
feature in the differential conductance. This feature moves further inside the gap for stronger coupling to the superconducting lead.
The position of the resonance depends only weakly on the normal lead coupling, which on the other
hand serves to broaden the resonance. In measurements, the sub-gap conductance peak could easily
be mistaken for a reduced-gap peak, and even more
so since at $eV=\Delta$ there is a dip instead of a peak. This
characteristic dip-peak structure has already been seen in
experiments in Refs.~\onlinecite{Deacon10b,Grove09}.
Another clear prediction resulting from the calculation is that
the position of the sub-gap structure moves away from the gap
edge, to lower voltages, when the gate potential is tuned away
from the particle-hole symmetric point in the middle of the
diamond.

The notion of spin-induced sub-gap resonances in the differential
conductance of S/QD/N junctions leaves some interesting questions unanswered. First
of all, it is not understood how the Yu-Shiba-Rusinov state
gradually changes into a Kondo resonance with increased coupling to the normal lead.
Secondly, the influence of an applied magnetic field is not clear. Naively one would expect
the sub-gap conductance peaks to split in a $B$-field. However, from the
classical spin model we know that a spin dependent cotunneling
model gives a very similar $IV$ curve and from this analogy one
expects little dependence on magnetic field, other than the
overall suppression of superconductivity, of course. Finally,
we mention the interesting problem of a spinful QD coupled to two
superconducting leads, where the interplay between spin-induced sub-gap resonances
and multiple Andreev reflections can be expected o give rise to unusual transport features.\cite{SandJespersen,Eichler}

\section{Acknowledgements}

B.M.A. acknowledges support from The Danish Council for Independent
Research $|$ Natural Sciences.

\end{document}